\documentclass[aps, preprint, amsmath,amssymb, superscriptaddress]{revtex4-1}

\usepackage{graphicx}
\usepackage{xcolor,color}

\begin{document}

\title{Collapse of superconductivity in cuprates via ultrafast quenching of phase coherence}
\author{F.\,Boschini}
\email[]{boschini@phas.ubc.ca}
\affiliation{Department of Physics $\&$ Astronomy, University of British Columbia, Vancouver, BC, Canada}
\affiliation{Quantum Matter Institute, University of British Columbia, Vancouver, BC, Canada}
\author{E.\,H.\,da Silva Neto}
\affiliation{Department of Physics $\&$ Astronomy, University of British Columbia, Vancouver, BC, Canada}
\affiliation{Quantum Matter Institute, University of British Columbia, Vancouver, BC, Canada}
\affiliation{Max Planck Institute for Solid State Research, Stuttgart, Germany}
\affiliation{Department of Physics, University of California, Davis, CA, USA}
\author{E.\,Razzoli}
\author{M.\,Zonno}
\affiliation{Department of Physics $\&$ Astronomy, University of British Columbia, Vancouver, BC, Canada}
\affiliation{Quantum Matter Institute, University of British Columbia, Vancouver, BC, Canada}
\author{S.\,Peli}
\affiliation{Department of Mathematics and Physics, Universit\`{a} Cattolica del Sacro Cuore, Brescia, Italy}
\affiliation{Interdisciplinary Laboratories for Advanced Materials Physics (ILAMP),Universit\`{a} Cattolica del Sacro Cuore, Brescia, Italy}
\author{R.\,P.\,Day}
\author{M.\,Michiardi}
\affiliation{Department of Physics $\&$ Astronomy, University of British Columbia, Vancouver, BC, Canada}
\affiliation{Quantum Matter Institute, University of British Columbia, Vancouver, BC, Canada}
\affiliation{Max Planck Institute for Chemical Physics of Solids, Dresden, Germany}
\author{M.\,Schneider}
\author{B.\,Zwartsenberg}
\author{P.\,Nigge}
\affiliation{Department of Physics $\&$ Astronomy, University of British Columbia, Vancouver, BC, Canada}
\affiliation{Quantum Matter Institute, University of British Columbia, Vancouver, BC, Canada}
\author{R.\,D.\,Zhong}
\affiliation{Condensed Matter Physics and Materials Science, Brookhaven National Laboratory, Upton, NY, USA}
\author{J.\,Schneeloch}
\affiliation{Condensed Matter Physics and Materials Science, Brookhaven National Laboratory, Upton, NY, USA}
\affiliation{Department of Physics $\&$ Astronomy, Stony Brook University, Stony Brook, NY, USA}
\author{G.\,D.\,Gu}
\affiliation{Condensed Matter Physics and Materials Science, Brookhaven National Laboratory, Upton, NY, USA}
\author{S.\,Zhdanovich}
\author{A.\,K.\,Mills}
\author{G.\,Levy}
\author{D.\,J.\,Jones}
\affiliation{Department of Physics $\&$ Astronomy, University of British Columbia, Vancouver, BC, Canada}
\affiliation{Quantum Matter Institute, University of British Columbia, Vancouver, BC, Canada}
\author{C.\,Giannetti}
\affiliation{Department of Mathematics and Physics, Universit\`{a} Cattolica del Sacro Cuore, Brescia, Italy}
\affiliation{Interdisciplinary Laboratories for Advanced Materials Physics (ILAMP),Universit\`{a} Cattolica del Sacro Cuore, Brescia, Italy}
\author{A.\,Damascelli}
\email[]{damascelli@physics.ubc.ca}
\affiliation{Department of Physics $\&$ Astronomy, University of British Columbia, Vancouver, BC, Canada}
\affiliation{Quantum Matter Institute, University of British Columbia, Vancouver, BC, Canada}

\newpage


\maketitle
\textbf{
The possibility of driving phase transitions in low-density condensates through the loss of phase coherence alone has far-reaching implications for the study of quantum phases of matter. This has inspired the development of tools to control and explore the collective properties of condensate phases via phase fluctuations. Electrically-gated oxide interfaces \cite{CavigliaLAOSTO,EmergentOxides}, ultracold Fermi atoms \cite{Review_ColdAtoms,GaeblerColdAtoms}, and cuprate superconductors \cite{EmeryKivelson,CorsonCoherenceNat}, which are characterized by an intrinsically small phase-stiffness, are paradigmatic examples where these tools are having a dramatic impact. Here we use light pulses shorter than the internal thermalization time to drive and probe the phase fragility of the Bi$_2$Sr$_2$CaCu$_2$O$_{8+\delta}$ cuprate superconductor, completely melting the superconducting condensate without affecting the pairing strength. The resulting ultrafast dynamics of phase fluctuations and charge excitations are captured and disentangled by time-resolved photoemission spectroscopy.
This work demonstrates the dominant role of phase coherence in the superconductor-to-normal state phase transition and offers a benchmark for non-equilibrium spectroscopic investigations of the cuprate phase diagram.}

The value of the critical temperature ($T_c$) in a superconducting material is controlled by the interplay of two distinct phenomena: the formation of electron pairs and the onset of macroscopic phase coherence. While the pairing energy ($E_p$) is generally controlled by the bosonic modes that mediate the electronic interactions  \cite{ReviewJohnstonDevereaux,ReviewKordyuk2010}, the macroscopic phase $\Theta$ depends on the stability of the condensate against fluctuations and inhomogeneities. The energy scale relevant for phase fluctuations can be expressed via the Ginzburg-Landau theory as $\hbar\Omega_{\Theta}$=$[\hbar^2n_S(0)a]/2m^*$, where $m^*$ is the effective mass of the pairs, $a$ is a characteristic length and $n_S(0)$ is the zero-temperature superfluid density. In conventional superconductors $E_p \ll \hbar\Omega_{\Theta}$ and therefore $T_c$ is determined solely by thermal charge excitations across the superconducting gap, which act to reduce the number of states available for the formation of the superconducting condensate. 

In cuprate superconductors, the scenario is much more complex since the small superfluid density pushes $\hbar\Omega_{\Theta}$ down to a value that is very close to the pairing energy \cite{EmeryKivelson}:
the low density of the quasi-2D condensate within the Cu-O planes depresses $\hbar\Omega_{\Theta}$ as low as $\approx$\,15 meV in bismuth-based copper oxides \cite{BenfattoPRB2001,EmeryKivelson}.
Several equilibrium measurements on underdoped cuprate superconductors have reported a non-zero pairing gap up to $T \approx 1.5 \times T_c$  \cite{NernstFluctuations,DiamagnetismFluctuations} even in the absence of macroscopic phase coherence. Upon heating for example, high-resolution angle-resolved photoemission (ARPES) experiments have shown pair-breaking scattering phenomena to emerge sharply at $T_c$ while the pairing gap is still open, suggesting a direct connection between pair-breaking and the onset of the phase fluctuations \cite{NatCommShin2015,TDOSNatPhys}. 
In the same temperature range, non-equilibrium optical and THz experiments have given evidence for picosecond dynamics dominated by phase fluctuations above $T_c$ \cite{CorsonCoherenceNat,3pulseReflFluctuations,PerfettiTHzFluctuations}.

The present work is motivated by the idea that a light pulse shorter than the internal thermalization time may be used to manipulate the density of phase fluctuations in a high-$T_c$ superconductor independent of the number of across-gap charge excitations. This would open the possibility of investigating a transient regime inaccessible at equilibrium, where both phase fluctuations and charge excitations are controlled by the same temperature and thus inherently locked. 
Here we demonstrate this concept in the underdoped Bi$_2$Sr$_2$CaCu$_2$O$_{8+\delta}$ (Bi2212) superconductor ($T_c \sim 82$\,K) \cite{STM_Bi2212_Gomes,DingGapBSCCO}. Time- and angle-resolved photoemission spectroscopy (TR-ARPES) is used to evaluate the electronic spectral function which encodes information regarding the pair-breaking dynamics. We demonstrate that the pair-breaking scattering rate $\Gamma_p$,  which is experimentally \cite{NatCommShin2015,TDOSNatPhys} and microscopically \cite{NormanSelfEnergy,FranzPRB1998,KwonDorseyFluctuations} associated with the scattering of phase fluctuations, is indeed decoupled from the dynamics of the pairing gap and across-gap charge excitations. At and above the critical fluence $F_C \approx$\,15 $\mu$J/cm$^2$ \cite{ScienceLanzara,Zhang2013,PRBsmallwoodGapDyn}, the increase of $\Gamma_p$ is such that superconductivity is suppressed. Quantitatively, we observe that the non-thermal melting of the condensate \cite{Kusar2008,Giannetti2009,ScienceLanzara,Zhang2013,PRBsmallwoodGapDyn,ReviewGiannetti} is achieved when $\Gamma_p\approx\hbar\Omega_{\Theta}$. 

TR-ARPES provides direct snapshots of the one-electron removal spectral function $A(\textbf{k},\omega)$ \cite{ReviewDamascelli} and its temporal evolution \cite{Sentef2013,Kemper2017} due to the perturbation by an ultrashort pump pulse. The spectral function $A(\textbf{k},\omega)$ depends on both the electron self-energy $\Sigma(\omega)$=$\Sigma'(\omega)$+$i\Sigma''(\omega)$ and the bare energy dispersion $\epsilon_{\textbf{k}}$:
\begin{equation} \label{EQ:1}
    A(\textbf{k},\omega)=-\frac{1}{\pi}\frac{\Sigma''(\omega)}{[\omega-\epsilon_{\textbf{k}}-\Sigma'(\omega)]^2+[\Sigma''(\omega)]^2} \hspace{4mm} .
\end{equation}
For a superconductor, $\Sigma(\omega)$ at the Fermi momentum $k=k_F$ can be approximated well by
\begin{equation}\label{EQ:2}
    \Sigma(\omega)=-i\Gamma_s+\frac{\Delta^2}{(\omega+i\Gamma_p)} \hspace{5mm},
\end{equation}
where $\Delta$ is the superconducting gap amplitude, $\Gamma_s$ the single-particle scattering rate and $\Gamma_p$ the pair-breaking scattering rate, as proposed in Ref.\,\cite{NormanSelfEnergy}. When the condensate is fully coherent, \emph{i.e.} for $T\ll T_c$ at equilibrium, the pair-breaking scattering rate $\Gamma_p$ is expected to vanish. This term may be interpreted as relating to the finite-lifetime of a Cooper pair as a result of scattering from phase fluctuations \cite{FranzPRB1998,KwonDorseyFluctuations,NormanSelfEnergy}. 

\begin{figure*}
\centering
\includegraphics[scale=0.6]{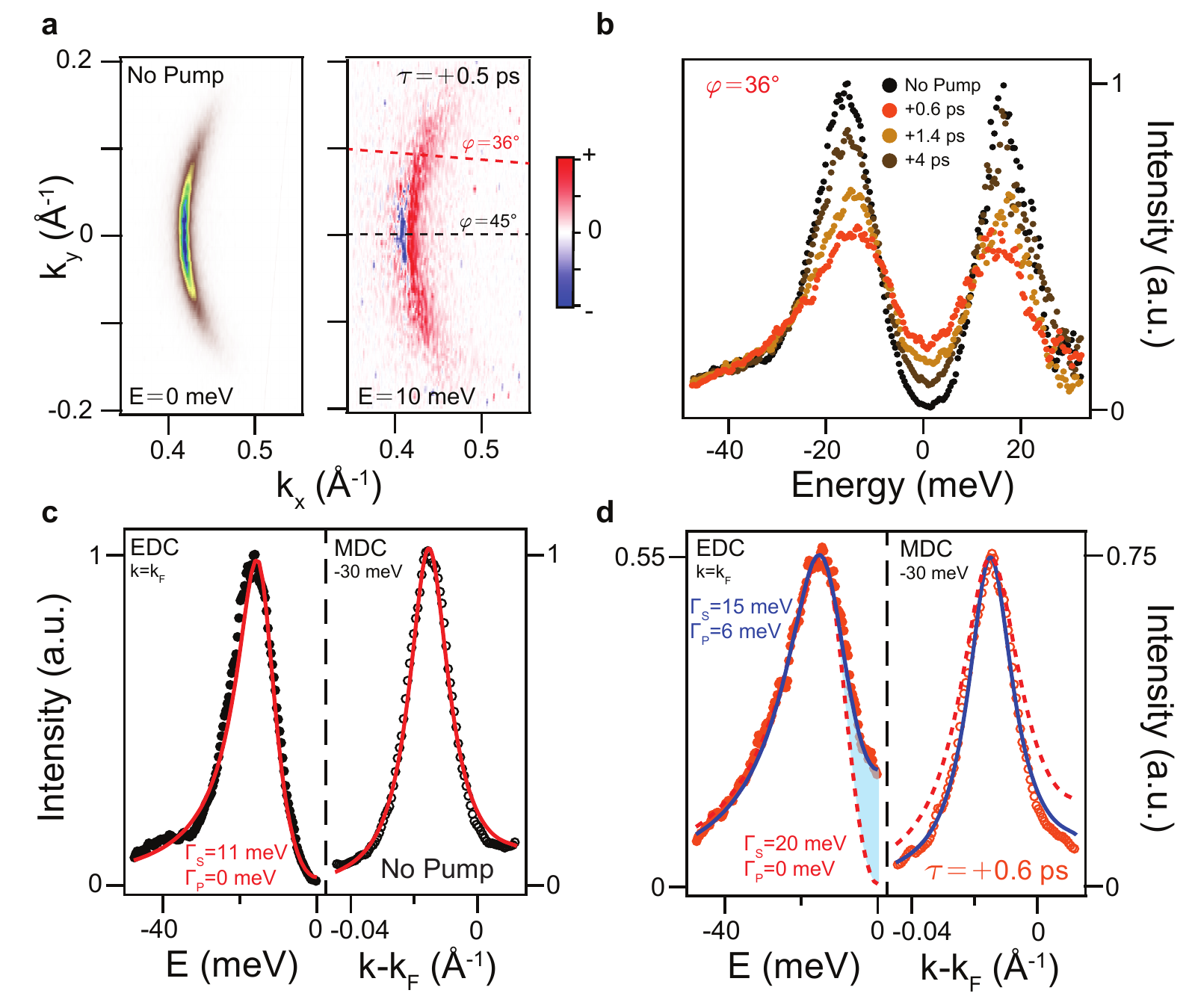}
\caption[Fig1]{\textbf{Ultrafast gap filling via enhancement of phase fluctuations.} \textbf{a} Equilibrium Fermi Surface mapping, left panel, and differential (Pump$_{\text{on}}$-Pump$_{\text{off}}$) iso-energy contour mapping at 10 meV above the Fermi level $E_F$, 0.5 ps pump-probe delay, right panel. The integration energy range is 10 meV and k$_{x}$ is aligned along the $\Gamma$-Y direction. The dashed black and red lines in the right panel define the nodal and off-nodal cuts investigated in the present work (details in Supplementary Information). 
\textbf{b} Off-nodal EDC at $k=k_F$ ($\varphi$=36$^o$) normalized to momentum-integrated nodal EDC ($\varphi$=45$^o$) at different pump-probe delays, $F<F_C$ fluence (F$_C\approx$15 $\mu J/cm^2$).
EDCs have been deconvoluted from the energy resolution broadening prior the division \cite{NatureSymm2008} (details in Supplementary Information).
\textbf{c} Equilibrium off-nodal ($\varphi$=36$^o$) normalized EDC at $k_F$ (left panel) and MDC at E=-30\,meV (right panel). The solid lines represent the best fit to the data. The EDC and MDC have been simultaneously fitted using a global procedure. Eqs.\,\ref{EQ:1}-\ref{EQ:2} are fitted to the EDC, while a phenomenological Lorentzian is fitted to the MDC deconvoluted from energy and angular resolutions, and additional contributions not accounted for due to the assumption of frequency independent scattering terms in Eq.\,\ref{EQ:2}. The equilibrium curve is well reproduced by $\Gamma_s$=11.0$\pm$0.5\,meV and $\Gamma_p\approx$0\,meV (red line).
\textbf{d} Non-equilibrium off-nodal ($\varphi$=36$^o$) EDC and MDC as measured at a delay of 0.6\,ps. The solid blue lines represent the outcome of the global fitting procedure, which gives $\Gamma_s$=15.0$\pm$0.5\,meV and $\Gamma_p$=6$\pm$1\,meV. The red dashed lines represent the curves obtained when $\Gamma_p$ is constrained to zero and $\Gamma_s$ is left as the only free parameter for the EDC fit non-benchmarked against the MDC. The transparent blue area highlights the filling of the superconducting gap induced by a sizable $\Gamma_p$.}
\label{Fig1}
\end{figure*}

To begin, we focus on the temporal evolution of the near-nodal superconducting gap. In Fig.\,\ref{Fig1}a we display a section of the Bi2212 Fermi surface (left panel) and the differential iso-energy contour map (right panel). The latter is obtained by subtracting the equilibrium iso-energy contour at $10$\,meV (above the Fermi energy, $E_F$) from its counterpart obtained at $0.5$\,ps pump-probe delay. This differential shows a clear in-gap signal that has been previously related to the quasiparticle (QP) recombination dynamics and the pairing gap closure \cite{ScienceLanzara,ShinTRARPESBSCCO, Zhang2016,Zhang2013,PRBsmallwoodGapDyn}. 
To then study the pairing gap dynamics, it is common to fit symmetrized energy distribution curves (SEDCs) at $k=k_F$ \cite{NormanSelfEnergy,NormanSymmetrizedEDCNature,ScienceLanzara}.
Although the emergence of a single peak in the SEDCs at large enough excitation fluences in TR-ARPES has been interpreted as a signature of the pump-induced gap closure \cite{ScienceLanzara,PRBsmallwoodGapDyn}, the comprehensive analysis of our data presented in the following provides clear evidence that a single peak in the SEDCs is instead related to the filling of an almost unperturbed pairing gap \cite{NatCommShin2015,TDOSNatPhys,DessauPRX}. This provides consistency between transient and equilibrium studies, offering a coherent picture of the electronic structure and its related dynamics. 

Before proceeding to the detailed modeling and quantitative analysis of the data, we address the microscopic origins of the evolution of the transient spectral function. We emphasize those photoinduced modifications to the spectral function which are immediately apparent, even at the level of visual inspection. In Fig.\,\ref{Fig1}b we present the temporal evolution of the low-fluence EDC at $k=k_F$ along the off-nodal direction ($\varphi$=36$^o$), normalized to the momentum-integrated nodal EDC (both deconvoluted from the energy resolution broadening prior the division, see section II of the Supplementary Information). Without invoking controversial symmetrization, this procedure allows us to explore the spectral function and its dynamics both below and above the superconducting gap. The resulting curves in Fig.\,\ref{Fig1}b provide direct evidence for the particle-hole symmetry of the quasiparticle states across the superconducting gap in the near-nodal region, \emph{i.e.} where pseudogap contributions are negligible \cite{BCScupratePRL,NatureSymm2008,ReviewGaps2014}. Most importantly, the data in Fig.\,\ref{Fig1}b reveal that the gap size (peak-to-peak distance) remains almost constant over the entire domain of time-delays measured. In contrast to this, we observe a transient decrease and broadening of the QP peak on either side of the gap, leading to a filling of spectral weight inside the superconducting gap (analogous conclusions are reached by a complementary analysis of the tomographic density of states \cite{TDOSNatPhys}, as shown in section III of the Supplementary Information).

For a more quantitative analysis, we can model the TR-ARPES data in terms of Eqs.\,\ref{EQ:1} and \ref{EQ:2}. In principle the in-gap broadening of the spectral function could be caused by both $\Gamma$ terms in Eq.\,\ref{EQ:2}, and so we have developed a global analysis of the EDCs and MDCs (momentum-distribution curves), which stabilizes the fitting procedure and achieves consistency across our results for all delays and excitation fluences (Supplementary Information Section IV). 
In Fig.\,\ref{Fig1}c we show the result of this global fitting at negative delays. The best simultaneous fit to EDC and MDC returns $\Gamma_s$=11.0$\pm$0.5\,meV and $\Gamma_p\approx$0\,meV, which are consistent with the equilibrium values extracted from conventional ARPES \cite{NatCommShin2015}. At positive delays (see the spectra at $\tau$=0.6\,ps in Fig.\,\ref{Fig1}d as a typical example), the filling of spectral weight inside the gap modifies the spectral lineshape such that even a qualitative fit requires the introduction of a non-zero $\Gamma_p$. Quantitatively, the sensitivity of the MDC lineshape to small variations of $\Gamma_s$ allows us to retrieve the values of the scattering rates at each time delay. This can be extended even as far as those excitations sufficiently large to induce the complete filling of the gap in spectral weight at $E_F$.

\begin{figure*}
\centering
\includegraphics[scale=0.75]{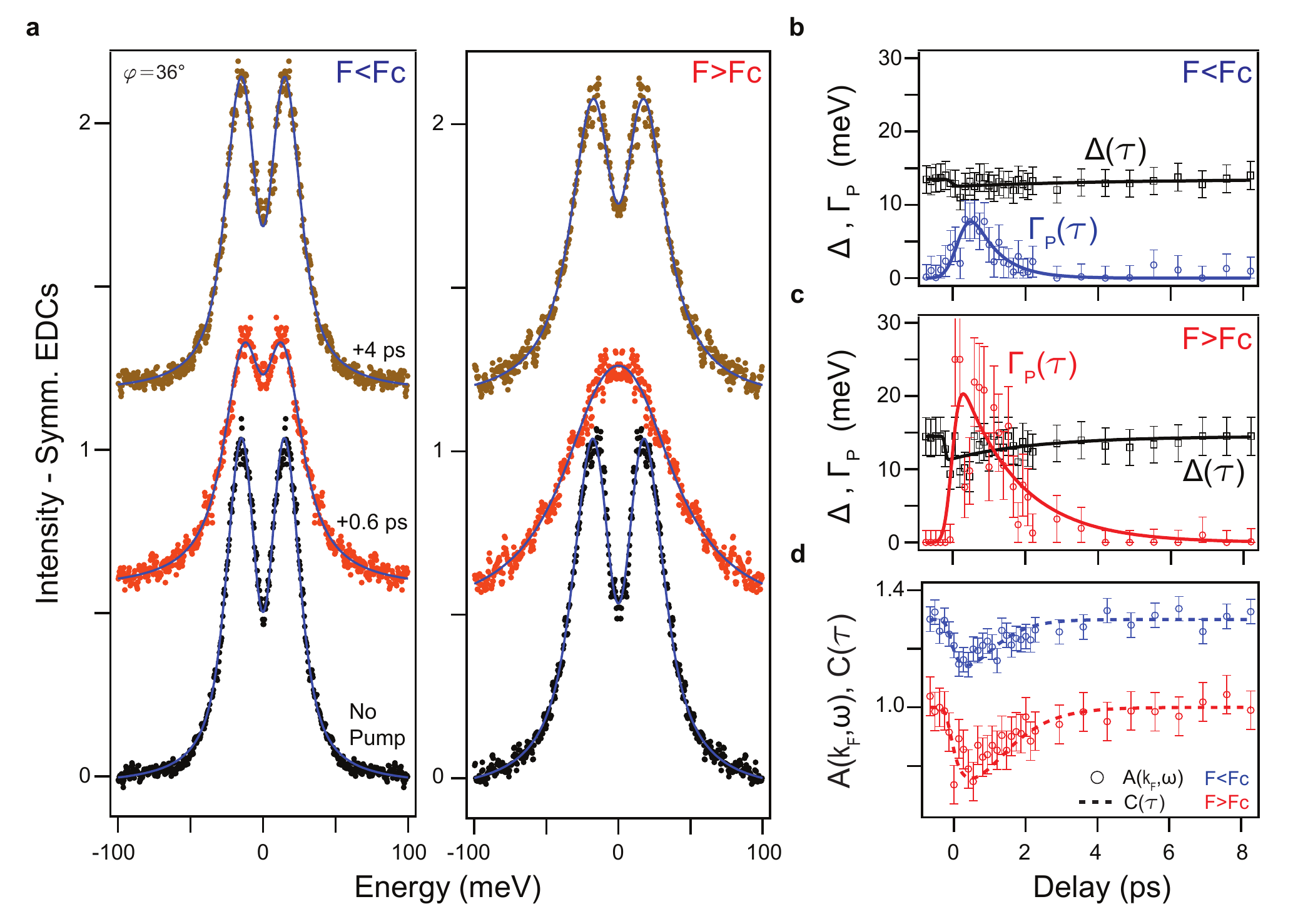}
\caption[Fig2]{\textbf{Temporal evolution of the spectral function via SEDC-MDC global analysis.} \textbf{a} Symmetrized EDCs (SEDCs) at $k=k_F$, off-nodal cut $\varphi$=36$^o$. SEDCs have been fitted using the global procedure described in the main text and Supplementary Information (blue lines).
\textbf{b}-\textbf{c} Ultrafast dynamics of $\Delta$ and the pair-breaking term, $\Gamma_p$, resulting from the global analysis of the SEDCs (shown in panel a) and MDCs. Pump excitation fluences are defined as F$<$F$_C$, panel b, and F$>$F$_C$, panel c. Solid lines are a phenomenological fit to a bi-exponential function convolved with a Gaussian accounting for the temporal resolution. 
\textbf{d} Temporal evolution of the amplitude of the spectral function at $k=k_F$ (normalized at $\tau<0$ ps, circles) and of the phenomenological function $C(\tau)$ as defined in the main text (dashed lines). Error bars in \textbf{b}-\textbf{d} define the confidence interval of the global procedure.
}
\label{Fig2}
\end{figure*}

We now move to the analysis of the temporal dynamics of $\Gamma_p$. For the sake of simplicity - and having experimentally verified particle-hole symmetry across the gap in the momentum range of interest - we analyze the SEDCs, which are not influenced by the effects of thermal broadening \cite{NormanSymmetrizedEDCNature} or the low signal to noise ratio for states above $E_F$, as in Fig.\,\ref{Fig1}.
In Fig.~\ref{Fig2}a we present the temporal evolution of the SEDCs along the off-nodal cut ($\varphi$=36$^o$) at two different excitation fluences, F$<$F$_C$ and F$>$F$_C$, where F$_C$ is the critical fluence for which the SEDCs exhibit a single envelope centered at the $E_F$ \cite{NodePeakLanzara,Zhang2013,PRBsmallwoodGapDyn}. 
For both fluences employed, the global fit approach described above provides an accurate and reliable determination of the temporal evolution of $\Gamma_s$ as well as $\Delta$ and $\Gamma_p$ (Fig.\,\ref{Fig2}b-c).
While the gap amplitude ($\Delta$) does not show a significant reduction for any excitation fluence, the leading term that drives the dynamics, and eventually the complete filling of spectral weight inside the gap, is the enhancement of $\Gamma_p$ as triggered by the pump excitation. 

As an interesting consequence, we note that the dynamics of the QP spectral weight (circles in Fig.\,\ref{Fig2}d) can be mapped onto the phenomenological function $C(\tau)=\frac{1}{2}[1+e^{-\frac{\Gamma_p(\tau)}{\Gamma_s(\tau)}}]$ (dashed lines in Fig.\,\ref{Fig2}d), which resembles the momentum-averaged two-particle scattering coherence factor \cite{CohFacQPI,HintonCohFac}. This empirical relationship between the single-particle ARPES spectral weight and a two-particle correlator suggests an intriguing scenario in which the QP peak amplitude is intertwined with the condensate density. Such a relationship, already suggested by previous ARPES studies \cite{FengSuperfluid,Ding_AN_superfluid,NodePeakLanzara}, calls for future experimental and theoretical investigations.

\begin{figure*}
\centering
\includegraphics[scale=0.55]{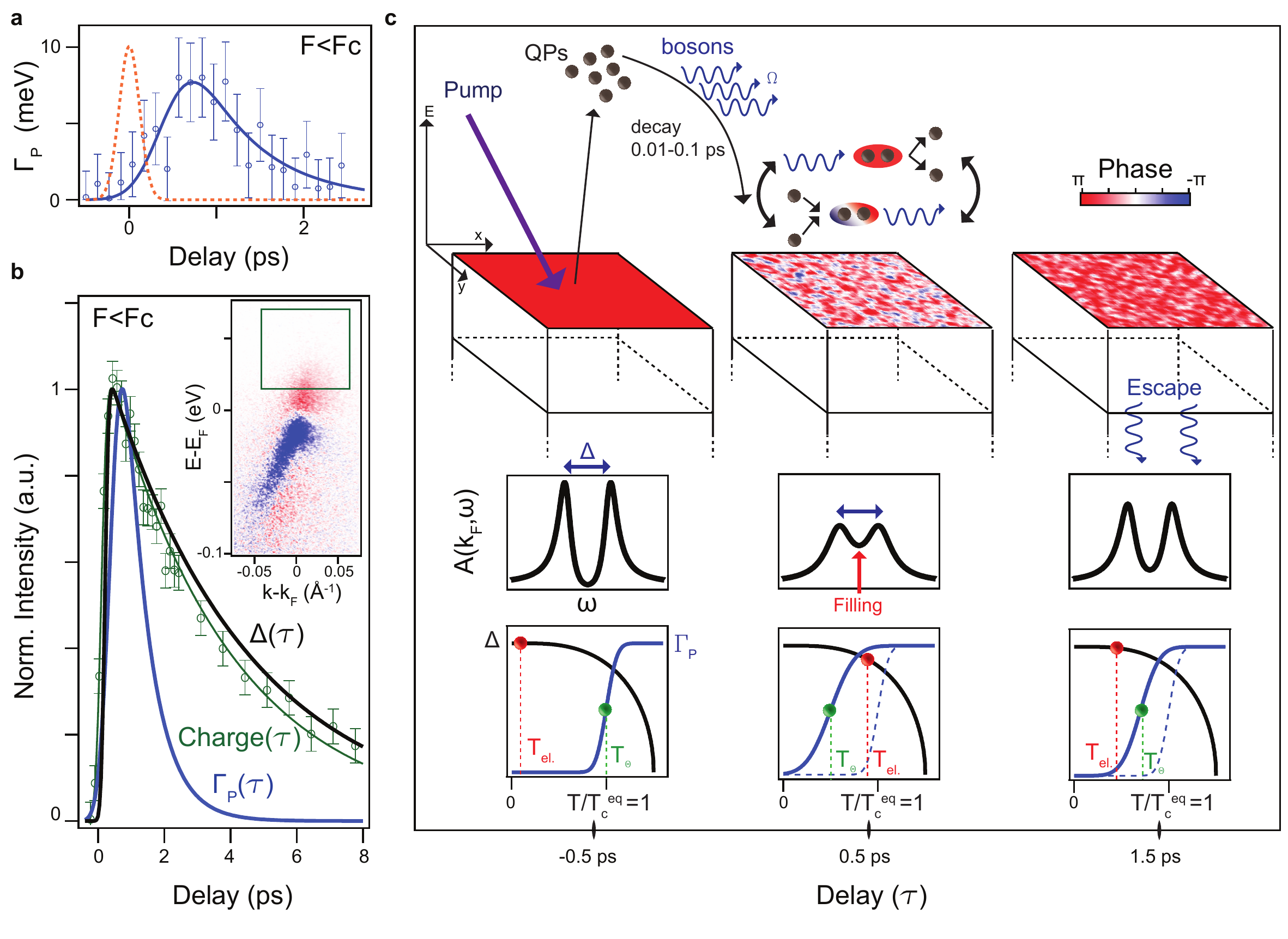}
\caption[Fig3]{\textbf{Role of phase fluctuations in the transient collapse of the condensate.} \textbf{a} $\Gamma_p$ dynamics for $F<F_C$ (blue circles, error bars defined in Fig.\,\ref{Fig2}; the blue line is the phenomenological fit described in Fig.~\ref{Fig2}), compared with the pump-probe cross-correlation (orange dashed line). \textbf{b} Comparison, again for $F<F_C$, of the normalized differential dynamics calculated as $[a(\tau)-a(\tau<0)]/\max[a(\tau)]$, for $\Gamma_p$ (blue line, from data in Fig.~\ref{Fig2}b), $\Delta$ (black line, from data in Fig.~\ref{Fig2}b), and charge dynamics (green circles obtained from the integrated off-nodal pump-induced population in the above-gap 15-70 meV energy window - highlighted area in the inset - and corresponding exponential fit; error bars represent the systematic errors associated with the experiment). \textbf{c} Pictorial sketch of the transient collapse of the condensate: a non-equilibrium bosonic population induces phase fluctuations leading to a gap filling and a modification of the temperature where the phase coherence is set independently to the charge dynamics. Top panels show a cartoon of the energetics of the process and the related real space condensate phase coherence; mid panels display the spectral function at $k=k_F$ when phase fluctuations are induced; 
bottom panels show the temporal evolution of the paring strength $\Delta$ (gap amplitude, black line) and of the pair-breaking scattering rate $\Gamma_p$ (blue line). While the pairing is controlled by the electronic temperature $T_{\text{el.}}$ (red spheres and dashed lines), and has an onset higher than T$_c$ itself, superconductivity and the macroscopic T$_c$ are determined by the onset of phase coherence at $T_\Theta \approx \hbar\Omega_{\Theta}/k_B$ (green spheres and dashed lines).} 
\label{Fig3}
\end{figure*}

The viability of measuring the evolution of $\Gamma_p$ in the time-domain provides essential information regarding the intrinsic dynamics of condensate formation in the cuprates.  Figure\,\ref{Fig3}a,b show that the $\Gamma_p$ relaxation dynamics for $F<F_C$ are completely decoupled from those of the gap amplitude and of the above-gap charge excitations. In particular, in Fig.\,\ref{Fig3}b we compare the temporal evolution of $\Gamma_p$ (blue, obtained by fitting the data in Fig.\,\ref{Fig2}b) with the dynamics of the superconducting gap (black, from Fig.\,\ref{Fig2}b) and of the charge excitations (green, as obtained by integrating the off-nodal pump-induced charge population in the above-gap 15-70 meV energy window shown in the inset of Fig.\,\ref{Fig3}b). While the temporal evolution of the above-gap excitations and that of the gap amplitude are locked to each other with a 4.0\,$\pm$0.5\,ps recovery time, $\Gamma_p$ relaxes much faster with a relaxation rate $\tau_{\Theta}\approx$\,1 ps. This value is of the same order of magnitude as the phase-correlation time extracted from high-frequency conductivity and related to the motion of topological defects \cite{CorsonCoherenceNat}. 

Microscopically, the transient increase of phase fluctuations can be rationalized as a cascade process triggered by the optical pump, which initially breaks the electronic pairs and promotes hot QPs to energies well above $E_F$. During their decay, the non-thermal QP population can either couple directly to phase excitations or scatter off high-energy bosonic excitations on a time scale of tens (spin fluctuations) to hundreds (optical phonons) of femtoseconds \cite{NatPhysDalConte,PRL2007PerfettiBi2212,NatCommBovensiepen2016}. The subsequent absorption of these bosons can subsequently break additional Cooper pairs. Furthermore, any pair recombination process must emit a gap-energy boson in order to satisfy energy conservation, as described by the Rothwarf-Taylor equations \cite{ReviewGiannetti}. As a result, after a few hundreds of femtoseconds the initial excitation is converted into a non-thermal bosonic population. We speculate that such highly energetic bosons, coupled to the fermionic bath, can interact even indirectly with the macroscopic condensate. These bosons can be considered as a possible source of the excess phase fluctuations which give a finite lifetime to the Cooper pairs. This picture is corroborated by the observation of a maximum change in $\Gamma_p$ (Fig.\,\ref{Fig3}a) approximately 500 fs after the pump excitation. Such a value is compatible with the build-up time observed via time-resolved optical spectroscopy and has been justified as the time necessary for the growth of the non-thermal gap-energy bosonic population \cite{ReviewGiannetti}. 

Together, these observations imply that the pair breaking processes related to the loss of coherence of the condensate can be decoupled from the charge excitations on the picosecond timescale. In this transient state, the condensate becomes more fragile, despite an almost unaffected pairing strength. This result has important consequences for establishing the nature of the instability of the macroscopic condensate at higher excitation fluences. Both time-resolved optical \cite{Averitt2001,THzProbeBSCCO,Kusar2008,Giannetti2009,ReviewGiannetti} and photoemission \cite{ScienceLanzara,NodePeakLanzara,Zhang2013,PRBsmallwoodGapDyn,ShinTRARPESBSCCO,PerfettiFillingGap,DessauPRX} experiments have measured the collapse of superconductivity and the complete quench of the coherence factor for pump fluence ranging from 14 to 70 $\mu$J/cm$^2$. Our data demonstrate that at $F\geq$\,15\,$\mu$J/cm$^2$  the non-equilibrium pair-breaking rate becomes of the order of the energy scale relevant to phase fluctuations, \emph{i.e.} $\Gamma_p\approx\hbar\Omega_{\Theta}\approx$\,15\,meV (Fig.\,\ref{Fig2}b), which corresponds to a Cooper pair lifetime of $\approx$ 40 fs. Figure\,\ref{Fig3}c provides a pictorial illustration of the dynamics of the superconductor-to-normal state phase transition: the transient excess of phase fluctuations driven by highly energetic bosons fills the superconducting gap and does not affect the pairing strength. We emphasize that while these results are consistent with the notion of preformed Cooper pairs and that sufficient enhancement of $\Gamma_p$ could culminate in the evolution of Fermi arcs \cite{NormanSelfEnergy,ChubukovGaplessSC}, the limited region of momentum space explored in this current work precludes any discussion of the pseudogap state. 

These results challenge the current understanding of the superconducting phase transition in cuprates. The TR-ARPES data presented here constitutes direct evidence that the phase coherence controls the condensate formation in underdoped high-$T_c$ superconductors, while the temperature-driven occupation of states plays a secondary role \cite{EmeryKivelson}. Indeed, our results demonstrate that the recovery of phase coherence is the primary and fastest mechanism by which we restore superconductivity (see Fig.\,\ref{Fig3}b). In addition, the ability to melt the condensate without altering the gap size or increasing the electronic temperature substantively (Fig.\,\ref{Fig3}c) suggests spectroscopic explorations of the hierarchy of pairing and phase coherence throughout the cuprate phase diagram, and in the vicinity of the putative quantum critical points \cite{sachdevQCP}. 
Further investigation and the development of selective excitation schemes will be essential to test possible interpretations of the dynamical response of the phase coherence in high-$T_c$ superconductors. In particular, a detailed study of the frequency dependence of $\Gamma_p$ may elucidate the microscopic mechanism responsible for the enhancement of phase fragility reported here. Furthermore, by extending these techniques to other members of the cuprate family, the relative role of dimensionality and interlayer coupling in the transient quenching of the superconducting condensate may be established \cite{CarboneBilayerCoherence,EmeryKivelson}.
The mechanism by which fermions interact with phase modes and how gap-energy bosons interact with the pair condensate toward the ultimate result of a plasma of incoherent excitations, still remains as an open and intriguing issue.\\

\section*{Methods}
\subsection*{Experimental set-up.}
Our TR-ARPES system is based on a Ti:Sapphire laser (VitesseDuo + RegA 9000 by Coherent) delivering 800 nm pulses (1.55 eV) with a 180-fs pulse duration, 250-kHz repetition rate. The output beam is split: a portion is used as the pump beam while the remaining part generates its fourth-harmonic, \emph{i.e.} 200 nm (6.2 eV). The 6.2-eV is generated through a cascade of nonlinear processes. The probe (6.2 eV) and the pump (1.55 eV) beams are both vertically (s) polarized and they are focused onto the sample (45$^o$ angle of incidence) using the same focusing optic leading to approximately 120 $\mu$m and 250 $\mu$m spot-sizes, respectively. 
The ARPES measurements are conducted in ultra-high-vacuum with a base pressure lower than 3$\cdot$10$^{-11}$ Torr, at a base temperature of 6 K. The angle and energy of the photoelectrons are resolved using a SPECS Phoibos 150 electron analyzer. The momentum, energy and temporal resolutions of the system are $<$0.003 $\text{\AA}^{-1}$, 19 meV and 250 fs, respectively, referenced from polycrystalline gold. Incident pump fluences indicated as F$<$F$_C$ and F$>$F$_C$ correspond to 8\,$\pm$\,2 $\mu$J/cm$^2$ and 30\,$\pm$\,4 $\mu$J/cm$^2$, respectively. 
\subsection*{Samples.} 
Single crystal Bi$_2$Sr$_2$CaCu$_2$O$_{8+\delta}$ (Bi2212) samples have been grown using the floating zone method and hole-doped by oxygen annealing (T$_c\simeq$82\,K). Bi2212 samples have been characterized by scanning tunneling microscopy measurements \cite{STM_Bi2212_Gomes}, and the gap amplitude extracted from the global fitting procedure agrees well with that reported elsewhere \cite{DingGapBSCCO}. \\

\begin{acknowledgments}
We thank L. Benfatto, A. Chubukov and M. Franz for useful and fruitful discussions.
C.G. acknowledge financial support from MIUR through the PRIN 2015 Programme (Prot. 2015C5SEJJ001) and from Universit\`{a} Cattolica del Sacro Cuore through D.1, D.2.2 and D.3.1 grants.
This research was undertaken thanks in part to funding from the Max Planck-UBC-UTokyo Centre for Quantum Materials and the Canada First Research Excellence Fund, Quantum Materials and Future Technologies Program. The work at UBC was supported by the Gordon and Betty Moore Foundation's  EPiQS Initiative, Grant GBMF4779, the Killam, Alfred P. Sloan, and Natural Sciences and Engineering Research Council of Canada’s (NSERC’s) Steacie Memorial Fellowships (A.D.), the Alexander von Humboldt Fellowship (A.D.), the Canada Research Chairs Program (A.D.), NSERC, Canada Foundation for Innovation (CFI), CIFAR Quantum Materials and CIFAR Global Scholars (E.H.d.S.N.). E.R. acknowledges support from the Swiss National Science Foundation (SNSF) grant no. P300P2-164649.
GDG is supported by the Office of Basic Energy Sciences, Division of Materials Sciences and Engineering, U.S. Department of Energy under contract No. DE-AC02-98CH10886. JS and RDZ  are supported by the  Center for Emergent Superconductivity, an Energy Frontier Research Center funded by the U.S. Department of Energy, Office of Science.  
\end{acknowledgments}

\hfill \break
\clearpage

\newpage
\newpage

\setcounter{figure}{0}
\renewcommand{\thefigure}{S\arabic{figure}}
\setcounter{equation}{0}
\renewcommand{\theequation}{S\arabic{equation}}

\part*{Supplementary Information}
\section{Equilibrium data analysis} \label{Equilibrium}
\begin{figure*}[b]
\centering
\includegraphics[scale=0.55]{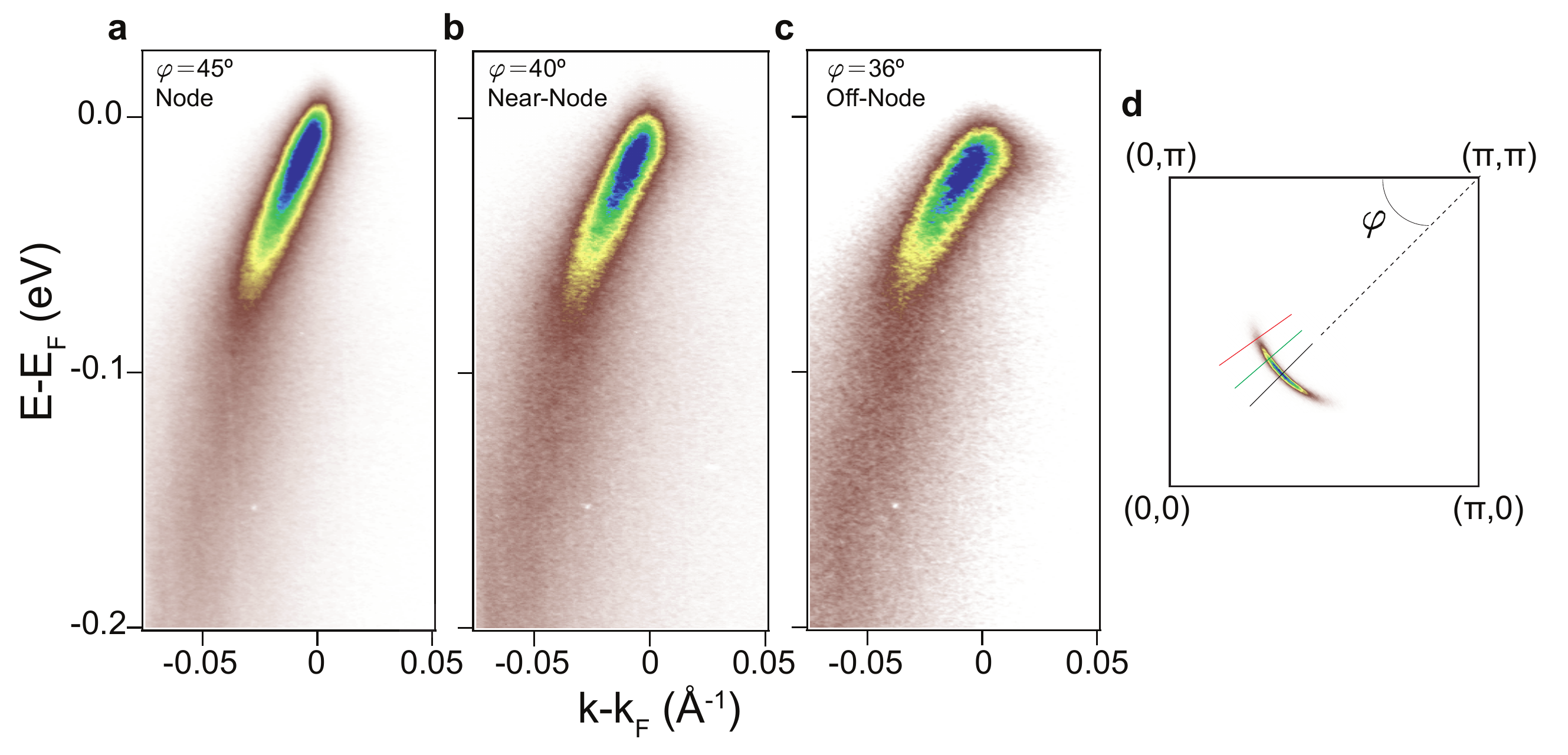}
\caption[FigS1]{\textbf{a-c} Band mapping at the equilibrium with s-polarized 6.2 eV light, base temperature 6 K, along the nodal ($\varphi = 45^o$, panel a), near-nodal ($\varphi = 40^o$, panel b) and off-nodal ($\varphi = 36^o$, panel c) directions, respectively. \textbf{d} Fermi surface mapping in a quadrant of the Brillouin zone. The sample was aligned along the $\Gamma - \text{Y}$ direction. The $\varphi$ angle is defined with respect to the (0,$\pi$)-($\pi$,$\pi$) direction. The three colored lines indicate the three measured momentum cuts: the nodal cut (black), a near-nodal cut (green) and an off-nodal cut (red).}
\label{Fig. S1}
\end{figure*}
As an experimental technique, angle-resolved photoemission spectroscopy (ARPES) offers an explicit connection to many-body theory via the association between photoemission intensity and the one-electron removal spectral function. From Fermi's Golden Rule, the ARPES intensity may be expressed as \cite{ReviewDamascelli}
\begin{equation} 
I(\textbf{k},\omega)=A(\textbf{k},\omega)\cdot|M|^2 \cdot f(\omega)\ast R(\omega), \label{PESintensity}
\end{equation}
where $A(\textbf{k},\omega)$ is the one-electron removal spectral function, $f(\omega)$ the electronic distribution (represented by the Fermi-Dirac distribution in equilibrium), $R(\omega)$ represents convolution with an experimental Gaussian resolution function, and $|M|^2$ the dipole matrix-element. As $|M|^2$ is largely dependent on light polarization and orbital character of the electronic states probed, it should be constant throughout our study and is neglected in the following. This expression may be further reduced to allow for direct consideration of the spectral function, and by extension the electronic self energy $\Sigma=\Sigma' +i\Sigma''$ (Eqs.\,1-2 in the main text). 

In order to do so, we note that Fermi statistics restrict static photoemission experiments to consideration of those states within $O(k_B T)$ of $E_F$ and below. However, in cases where the spectral function is particle-hole symmetric at the Fermi-momentum $k = k_F$ \cite{BCScupratePRL}, the ARPES spectra $I(\textbf{k},\omega)$ can be symmetrized about $E=E_F$ to overcome this limitation. The resulting symmetrized energy distribution curves (SEDCs) are then independent of $f(\omega)$ \cite{NormanSymmetrizedEDCNature}. Combined then with our assumption of constant matrix elements, SEDC$(\omega)\propto A(k_F,\omega) \ast R(\omega)$.
\begin{table}[t]
\caption{\label{table1}Parameters extracted from the fitting procedure shown in Figure ~\ref{Fig. S2}.}
\begin{tabular}{||c | c | c | c ||}
\hline
\hline
Cut $\varphi$ (deg) & $\Delta$ (meV) & $\Gamma_s$ (meV) & $\Gamma_p$ (meV) \\
\hline
\textbf{45} & 0 & 10.9 $\pm$ 0.2 & 0\\
\hline
\textcolor{green}{\textbf{40}} & 6 $\pm$ 0.4 & 10.2 $\pm$ 0.4 & 0\\
\hline
\textcolor{red}{\textbf{36}} & 14.7 $\pm$ 0.2 & 11.8 $\pm$ 0.2 & 0\\
\hline
\hline
\end{tabular}
\end{table} 
\begin{figure*}[b]
\centering
\includegraphics[scale=0.67]{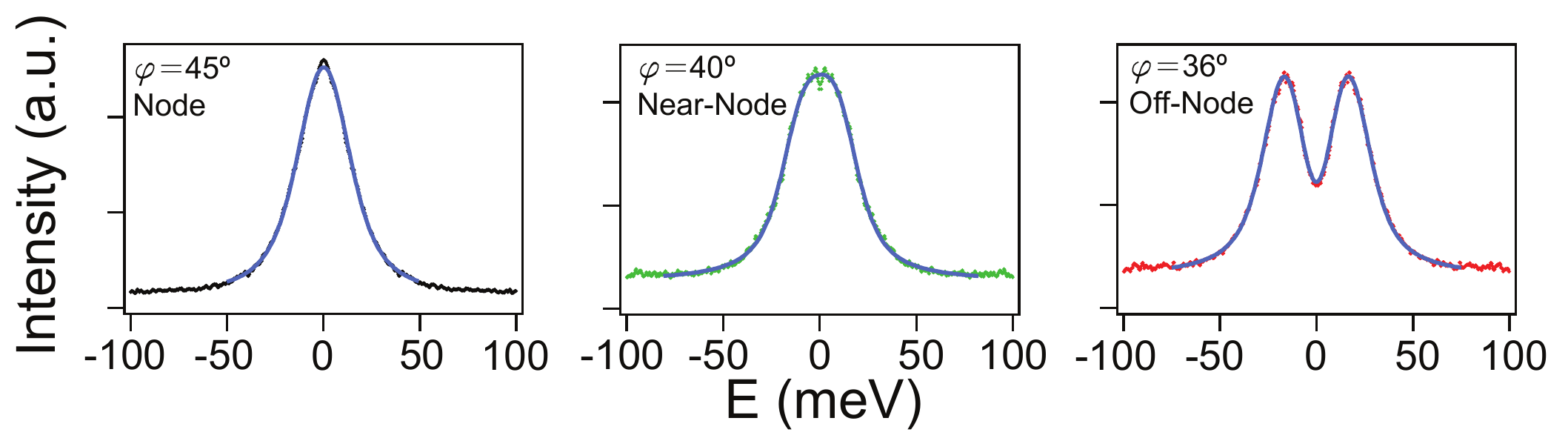}
\caption[FigS2]{Equilibrium SEDCs for the three measured momentum cuts. The blue lines show the fit resulting from Eqs.~1-2 in the main text.}
\label{Fig. S2}
\end{figure*}
With this interpretation of the ARPES intensity in mind, we present here static ARPES data, and SEDCs at $k_F$ to establish the equilibrium system under consideration. Figure~\ref{Fig. S1} displays raw data collected at T = 6 K with 6.2-eV s-polarized light. In Fig.~\ref{Fig. S1}a-c, the ARPES intensity along the nodal ($\varphi = 45^o$), near-nodal ($\varphi = 40^o$) and off-nodal ($\varphi = 36^o$) directions may be compared. The three cuts along nodal (black), near-nodal (green) and off-nodal (red) are illustrated in relation to their position within the Brillouin zone by the constant-binding energy map ($E=E_F\pm7.5$ meV) in Fig.~\ref{Fig. S1}d.
In Fig.~\ref{Fig. S2}, equilibrium SEDCs along the three momentum cuts in Fig.~\ref{Fig. S1} are plotted alongside fits to Eqs.\,1-2 of the main text. The extracted fit parameters are presented in Table~\ref{table1}. In agreement with Kondo et al. \cite{NatCommShin2015}, $\Gamma_p$=0 at T$\ll$T$_C$ while $\Gamma_s$ is finite and nearly momentum-independent in the near-nodal region. The result is further validated by the agreement between the gap parameters extracted from this fitting and those reported elsewhere \cite{DingGapBSCCO}.

\section{Direct access to the transient spectral function} \label{deconvolution}
\begin{figure*}[b]
\centering
\includegraphics[scale=0.55]{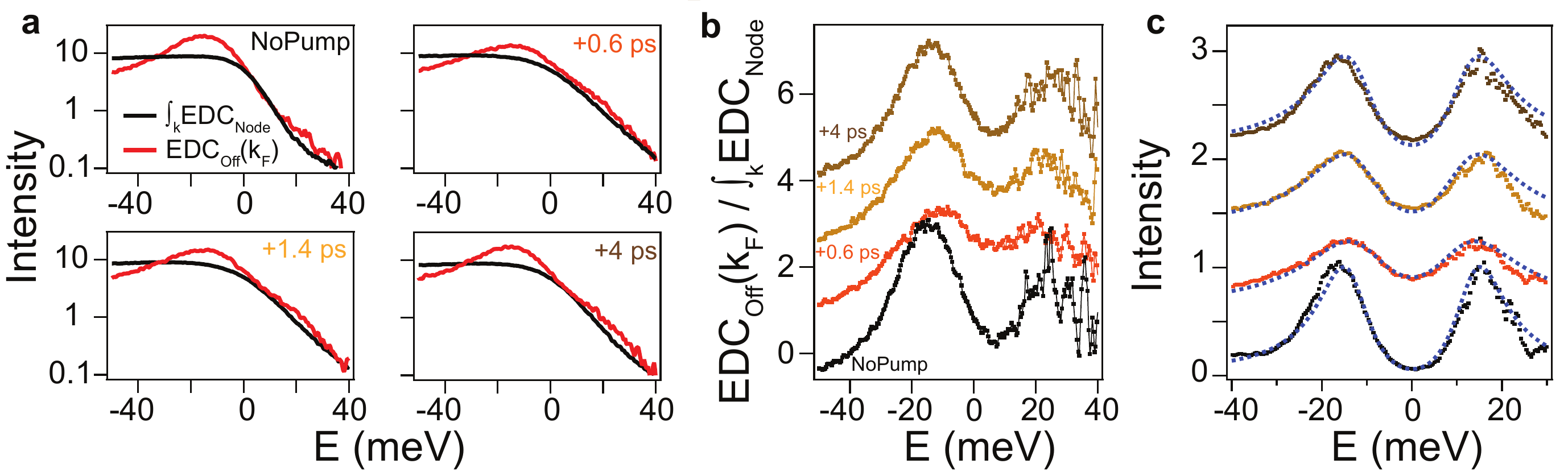}
\caption[EDCdivFD]{\textbf{a} Momentum-integrated EDCs along the nodal cut (black line) and EDC at $k=k_F$ along the off-nodal cut ($\varphi$=36$^o$) at different pump-probe delays. \textbf{b} Off-nodal EDCs normalized to momentum-integrated nodal EDCs at different delays. \textbf{c} Same as panel b but after deconvolving the EDCs from the energy resolution using the Lucy-Richardson deconvolution procedure. The blue dashed lines show the spectral function calculated using parameters shown in Figure \,2b-c}
\label{EDCdivFD}
\end{figure*}

Characterization of the transient spectral function both below and above $E_F$ is complicated by the suppression of photoemission intensity above $E_F$ due to the nature of the electronic distribution $f(\omega)$. We seek then to establish a method by which to address the spectral function above $E_F$ directly, minimizing the number of assumptions made in this process.
This may be achieved by establishing an experimental measure of $f(\omega)$ which we can then divide from the experimental intensity so as to access and amplify the intensity above $E_F$.
By integrating intensity along the nodal direction, the resulting momentum-integrated EDC ${\int I_{\text{node}}(k,\omega)dk}$ can be used to this purpose as it resembles the electronic distribution [$\int A_{\text{node}}(k,\omega)f(\omega)dk \propto f(\omega)$]. The nodal momentum-integrated EDCs are plotted (black curves) alongside off-nodal EDCs at $k=k_F$ (red curves) for various pump-probe delays and pump fluence $F<F_C$ in Fig.~\ref{EDCdivFD}a. 

Upon division by the nodal momentum-integrated EDCs, the double peak feature underlying the raw EDCs become apparent, as illustrated in Fig.~\ref{EDCdivFD}b. As a consequence of convolution with the energy resolution function $R(\omega)$, the peaks are not symmetric about $E=0$ meV. If however we deconvolve the nodal and off-nodal intensities from the energy resolution prior to the division we can overcome this limitation. To do so, we employ the Lucy-Richardson deconvolution procedure \cite{NatureSymm2008} (see Figure \,1b-d in the main text and Fig.~\ref{EDCdivFD}c). Following deconvolution, the ratio of the off-nodal EDC and momentum-integrated nodal EDC is proportional to the off-nodal spectral function
\begin{equation} 
\text{Ratio}=\frac{I_{\text{off}}(k_F,\omega)}{\int I_{\text{node}}(k,\omega)dk}\propto A_{\text{off}}(k_F,\omega).\\ \label{RatioFun}
\end{equation}

This same approach is behind the implementation of the tomographic density of states method \cite{TDOSNatPhys}, as discussed in section III. The resulting curves demonstrate the particle-hole symmetry of the superconducting spectral function in the near-nodal region, in agreement with other ARPES studies \cite{NatureSymm2008,BCScupratePRL}, and act to validate the symmetrization technique applied to equilibrium and out-of-equilibrium ARPES. Blue dashed lines in Fig.~\ref{EDCdivFD}c display the spectral function calculated with the parameters extracted from the SEDC-MDC global fitting procedure (Figure \,1c-d and Figure \,2b-c, section IV) showing a remarkable one-to-one correspondence between these renormalized off-nodal EDCs and the SEDC-MDC global fitting procedure.

\section{Transient Tomographic Density of States} \label{TDOS}
\begin{figure*}[t]
\centering
\includegraphics[scale=0.5]{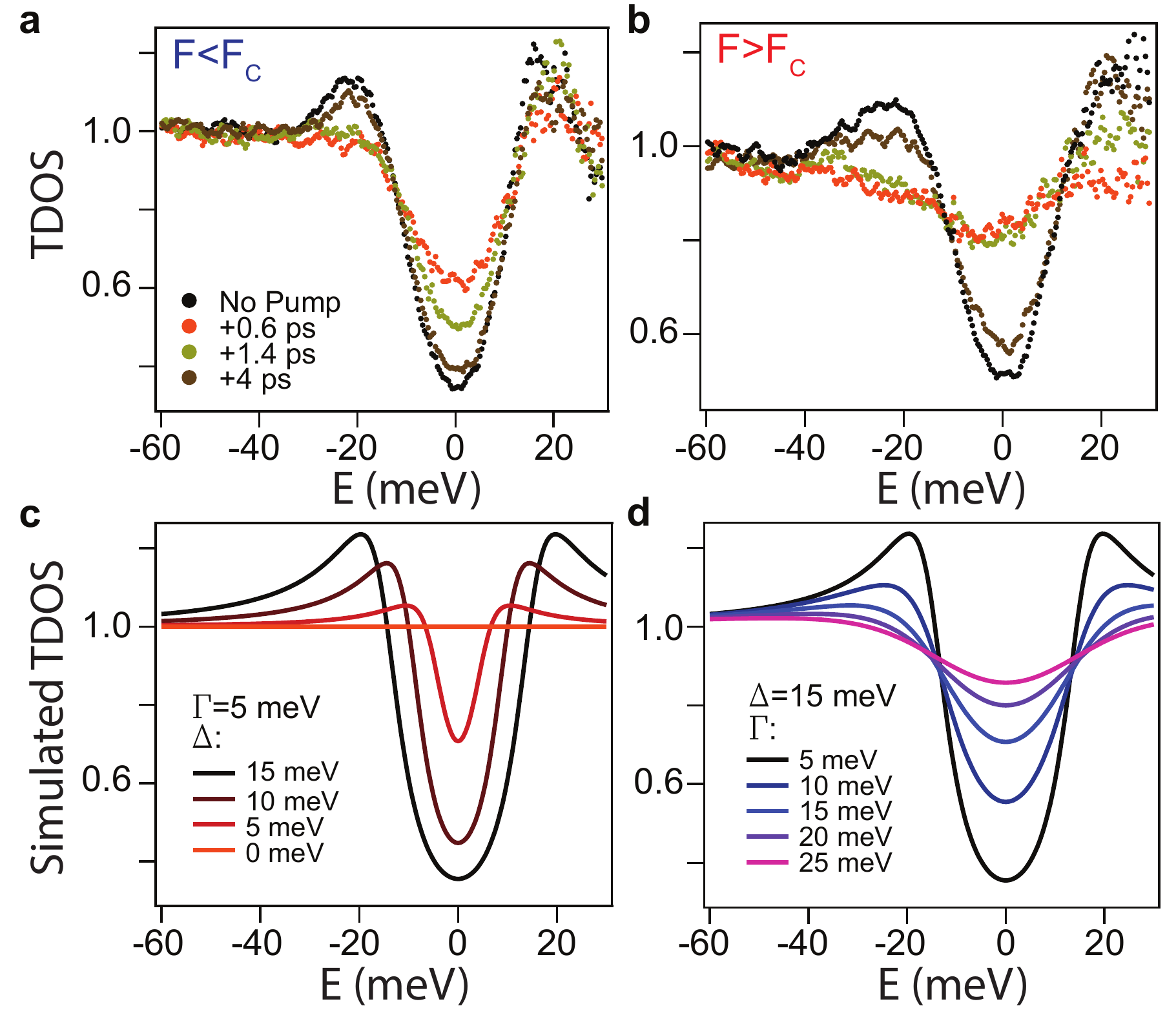}
\caption[TDOSvsSpectral]{\textbf{a-b} Transient TDOS for two different excitation fluences: F$<$F$_C$ and F$>$F$_C$. Momentum-integrated EDCs have been deconvoluted from the energy resolution (see section II). \textbf{c-d} Simulated TDOS curves using the Dynes function varying the gap amplitude (panel c) or the $\Gamma$ term (panel d).}
\label{TDOSvsSpectral}
\end{figure*}
In addition to the analysis of EDCs presented in the main text, the transient tomographic density of states (TDOS) provides a complementary proof of the gap filling. The TDOS is obtained as the ratio between the off-nodal and nodal EDCs integrated along a momentum cut perpendicular to the Fermi surface. As the primary difference between these EDCs is associated with the superconducting gap, this method allows for us to isolate consideration of the gap from other spectral features present in the EDCs. This confirms the most significant effect of the pump excitation to be a filling of the superconducting gap \cite{TDOSNatPhys}. 
The TDOS is defined as:
\begin{equation}
\text{TDOS}(\omega)=\frac{\int A_{\text{off}}(k,\omega)f(\omega)dk}{\int A_{\text{node}}(k,\omega)f(\omega)dk}.
\label{TDOS}
\end{equation}

In Fig.~\ref{TDOSvsSpectral}a-b we display the transient experimental TDOS for two different excitation fluences. The energy resolution has been deconvolved from the momentum-integrated EDCs \cite{NatureSymm2008}, as explained in section\,\ref{deconvolution}.
To illustrate the influence of gap closing and filling on the TDOS, we have simulated TDOS curves for a Dynes function \cite{TDOSNatPhys} 
\begin{equation}
\text{Dynes}(\omega)=\text{Re}\frac{\omega-i\Gamma}{\sqrt{(\omega-i\Gamma)^2-\Delta^2}},
\label{Dynes}
\end{equation}
for the case where $\Delta \rightarrow$\,0\,meV as for a gap closure and $\Gamma >$\,5\,meV for a gap filling. The result is plotted in Fig.~\ref{TDOSvsSpectral}c-d.
We note here that the single $\Gamma$ term in the Dynes function subsumes the $\Gamma_{s,p}$ terms from Eq.\,2 of the main text \cite{NatCommShin2015}. Our experimental observations (Fig.\,\ref{TDOSvsSpectral}a-b) are in agreement with the gap filling picture (Fig.~\ref{TDOSvsSpectral}d), for both high and low fluences. 

\section{SEDC-MDC Global Fit} \label{GlobalFit}
\begin{figure*}[t]
\centering
\includegraphics[scale=0.5]{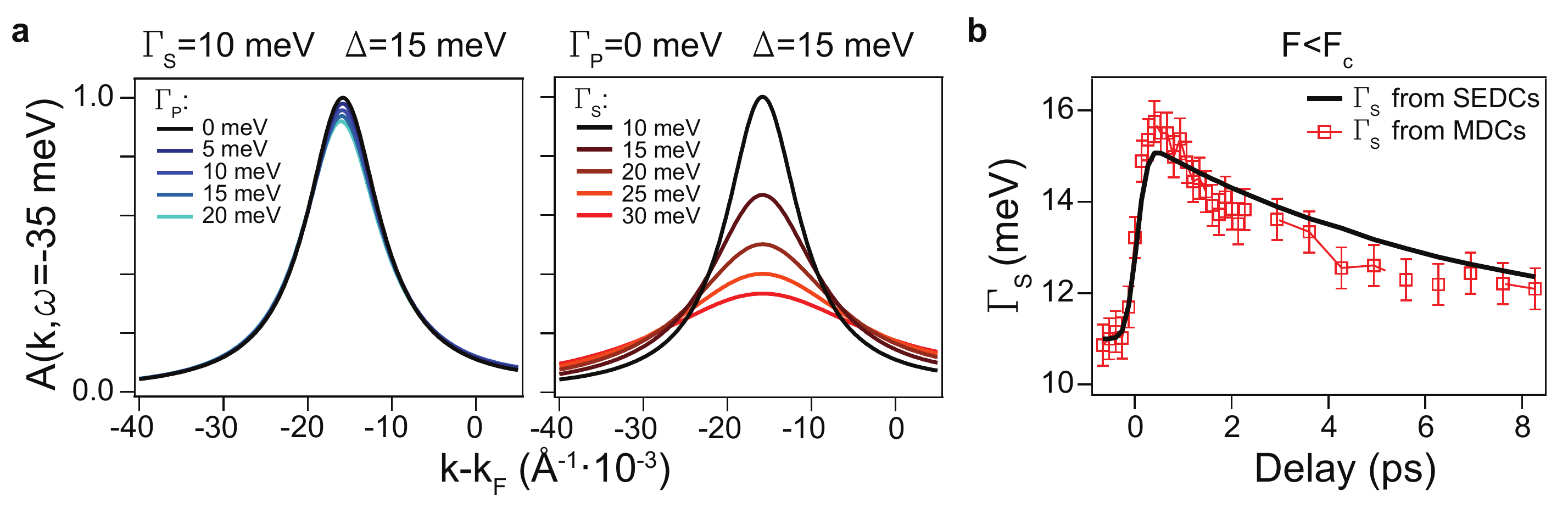}
\caption[FigScompA]{\textbf{a} Spectral function  $A(k,\omega$=-35 meV), related to MDCs, dependence on $\Gamma_p$ and $\Gamma_s$. \textbf{b} Temporal evolution of $\Gamma_s$ extracted by fitting SEDCs at F$<$F$_C$ (black line) compared to the one extracted from MDCs (red markers).}
\label{FigScompA}
\end{figure*}

As discussed in the main text, both $\Gamma_s$ and $\Gamma_p$ terms in Eq.~2 can act to broaden the in-gap spectral function. However, the effects of the two on the lineshape are indeed distinguishable. By addressing momentum-distribution curves (MDCs) rather than EDCs, we observe that the MDC width is primarily influenced by the $\Gamma_s$ term, and shown in Fig.~\ref{FigScompA} to be fairly insensitive to variations in $\Gamma_p$. To reliably extract the evolution of $\Delta$ and $\Gamma_p$ in different fluence regimes, we have therefore developed a global analysis of SEDCs and MDCs. The evolution of the MDC width as a function of pump-probe delay ($\tau$) can be used to then constrain and confirm the $\Gamma_s$ extracted from the SEDCs where $\Gamma_p$ and $\Gamma_s$ both contribute to the linewidth.

While the imaginary part of the electron self-energy is closely related to the MDC width $\Sigma''_{\text{MDC}}$ \cite{ReviewDamascelli}, the latter is in general larger than $\Gamma_s$ as a result of energy and angular resolutions, as well as additional frequency dependent contributions to the scattering terms not accounted for in Eq.~2 for simplicity. 
To estimate the extent of this additional broadening to the MDC width, we can compare the equilibrium ($\tau<0$) $\Sigma''_{\text{MDC}}$ with the $\Gamma_s$ as extracted from the SEDCs (section\,\ref{Equilibrium}). We assume here that the overall broadening of the MDCs will be unaffected by the pump excitation (see Fig.\,1c-d). Ultimately, we find good agreement between the temporal dynamics of $\Gamma_s$ as extracted from nodal SEDCs (black line in Fig.\,\ref{FigScompA}b) with the MDC widths (red squares in Fig.\,\ref{FigScompA}b). This confirms the SEDC fits to $\Gamma_s$ as well as placing firm upper limits on contributions to the self energy beyond Eq.~2. 

\section{Transient electronic temperature} \label{Te}
\begin{figure*}[t]
\centering
\includegraphics[scale=0.6]{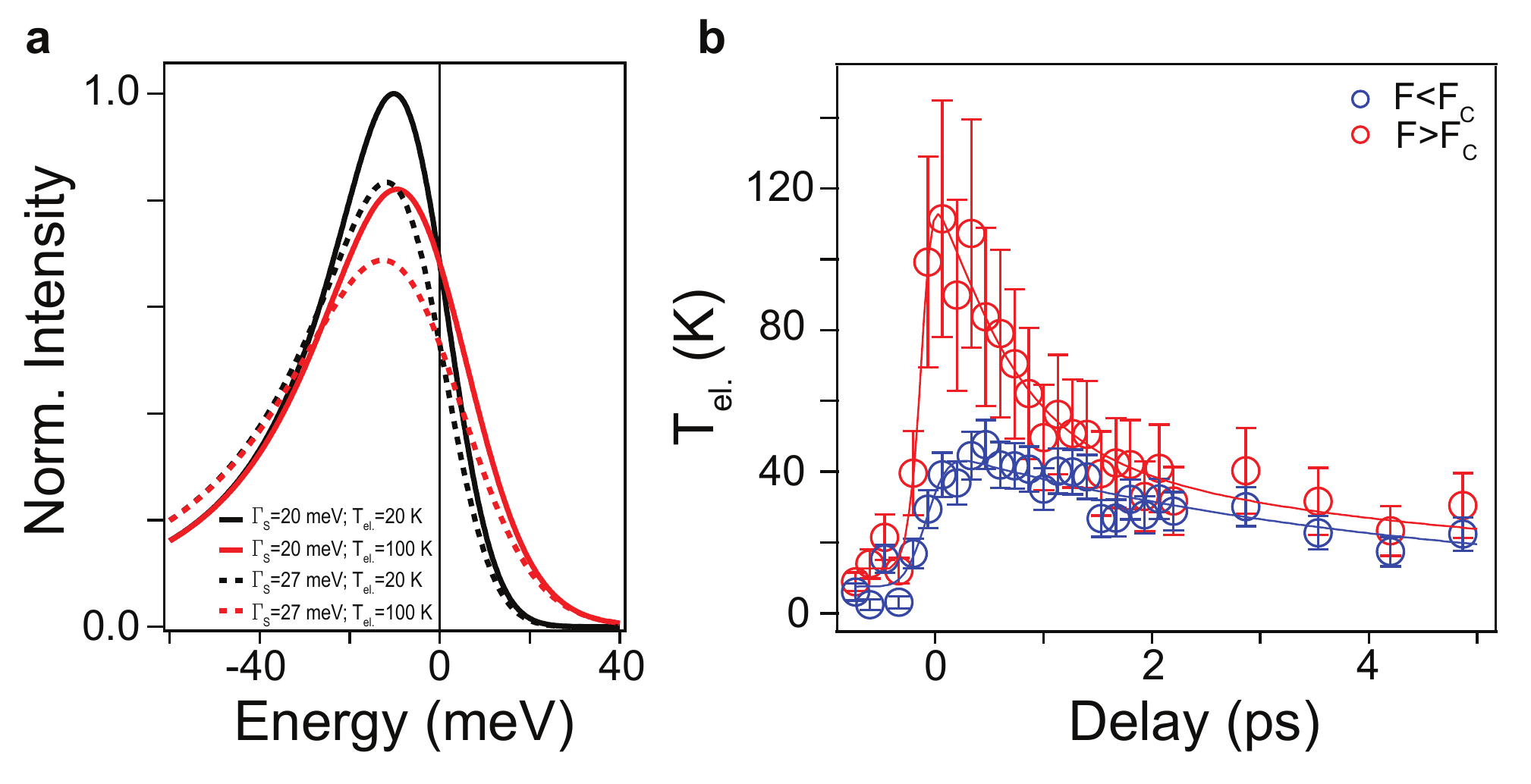}
\caption[FigS6]{\textbf{a} Simulated EDCs at $k=k_F$ along the nodal direction as function of $\Gamma_s$ and $T_{el.}$ parameters. EDCs have been convoluted with a 19 meV energy resolution. \textbf{b} Extracted dynamics of the electronic temperature $T_{el.}$ for both $F<F_C$ and $F>F_C$.}
\label{Fig. S6}
\end{figure*}
As the pump excitation will modify the electronic temperature, it is essential that we establish the pair-breaking scattering phenomena to originate from physics more substantive than a pure thermal effect brought on by the pump light. To do so, we extract an approximate transient electronic temperature $T_{el.}$ by fitting the EDC at $k=k_F$ along the nodal direction under various values of $\tau$ and different pump fluences. The fitting function $\text{Fit}(\omega)$ is defined as the product between the nodal spectral function at $k=k_F$ with amplitude $A_{\text{QP}}$ and the Fermi Dirac distribution $f_{\text{FD}}(\omega)=[e^{\frac{\omega}{k_B T_{el.}}}+1]^{-1}$ and it can be expressed as:
\begin{equation}
\text{Fit}(\omega)=[\frac{A_{\text{QP}}}{\pi}\frac{\Gamma_s}{\omega^2+\Gamma_s^2}\cdot f_{FD}(T_{el.}, \omega)] \ast R(\omega). \label{eqNode}
\end{equation}
Evidently, both $\Gamma_s$ and $T_{el.}$ will influence the nodal EDC lineshape. However, as with the $\Gamma_{s,p}$ terms in the MDC fits of the previous section, the influence of $T_{el.}$ and $\Gamma_s$ can be disentangled here. This is illustrated in Fig.~\ref{Fig. S6}a, where we show simulated nodal EDCs using Eq.~\ref{eqNode} by changing $\Gamma_s$ and $T_{el.}$ parameters. Evidently, if the temporal evolution of $\Gamma_s$ (see Fig.~\ref{FigScompA}b) was not included in the electronic temperature extraction procedure, $T_{el.}$ would be overestimated.
In Fig.~\ref{Fig. S6}b we show the extracted temporal evolution of T$_{el.}$ for both the employed fluences. While for F$>$F$_C$ fluence the maximum electronic temperature is $T_{el.}^{max}>$T$_C$, for F$<$F$_C$ $T_{el.}^{max}\approx$\,T$_C$/2. This provides additional evidence that the transient $\Gamma_p$ is not a simple thermal effect \cite{NatCommShin2015} but has a deeper physical meaning. In addition, static ARPES experiments \cite{NatCommShin2015,TDOSNatPhys} have shown that the superconducting gap amplitude follows a BCS-like temperature dependence with a closing temperature $T_{close}\approx$ 140 K. Thus, even for the F$>$F$_C$ fluence, where $T_e^{max}\approx$ 110 K\,$>T_C$, we expect a gap quenching of around 30$\%$, consistent with what has been reported in the main text (Fig.\,2c). \\

\bibliographystyle{apsrev4-1}
\providecommand{\noopsort}[1]{}\providecommand{\singleletter}[1]{#1}

\end{document}